\documentclass{osa-article}

\journal{oe}

\articletype{Research Article}

\usepackage{lineno}

\begin{document}

\title{A detail-enhanced sampling strategy in Hadamard single-pixel imaging}

\author{Yan Cai,\authormark{1} Shi-Jian Li,\authormark{1} Wei Zhang,\authormark{1} Hao Wu,\authormark{1} Xu-Ri Yao,\authormark{1,2,3} and Qing Zhao\authormark{1,2,4}}

\address{\authormark{1}Center for Quantum Technology Research and Key Laboratory of Advanced Optoelectronic Quantum Architecture and Measurements (MOE), School of Physics, Beijing Institute of Technology, Beijing 100081, China \\
\authormark{2}Beijing Academy of Quantum Information Sciences, Beijing 100193, China \\
\authormark{3}yaoxuri@bit.edu.cn \\
\authormark{4}qzhaoyuping@bit.edu.cn}

\begin{abstract*}
Hadamard single-pixel imaging (HSI) is an appealing imaging technique due to its features of low hardware complexity and industrial cost. To improve imaging efficiency, many studies have focused on sorting Hadamard patterns to obtain reliable reconstructed images with very few samples. In this study, we present an efficient HSI imaging method that employs an exponential probability function to sample Hadamard spectra along a direction with better energy concentration for obtaining Hadamard patterns. We also propose an XY order to further optimize the pattern-selection method with extremely fast Hadamard order generation while retaining the original performance. We used the compressed sensing algorithm for image reconstruction. The simulation and experimental results show that these pattern-selection method reliably reconstructs objects and preserves the edge and details of images. 
\end{abstract*}

\section{Introduction}
Single-pixel imaging (SPI) \cite{Gibson:20} is one of the most representative applications of compressed sensing (CS) theory \cite{Candes:08,Duarte:08,Donoho:06} in imaging. A SPI system uses a spatial light modulator to illuminate an object with structured light or modulate the light carrying the object information. A single-pixel detector is used to collect the partially sampled information light, which is then inputted into a reconstruction algorithm to recover an accurate image. SPI has the advantages of strong robustness, high sensitivity, and a wide-spectrum range \cite{Duarte:08}, it has been widely used in three-dimensional imaging \cite{SUN:13,Zhang:16,Sun:16}, terahertz imaging\cite{Chan:08,StantchevSun:16,Stantchev:17}, infrared imaging\cite{Johnson:19,Radwell:14}, fluorescence microscopy\cite{Liu:18,Chowdhury:15,Studer:12}, information encryption\cite{Zhang:19,Zhang:18,Chen:14,Ye:19}, and Lidar\cite{Pawlikowska:17,Kuzmenko:20,Tobin:19}, among others.

In SPI, the illumination pattern is a factor affecting the performance of the imaging system. SPI uses random matrices, such as Gaussian and Bernoulli matrices\cite{Candes:06}, as patterns to recover the image. Recently, the Hadamard single-pixel imaging (HSI) and Fourier single-pixel imaging (FSI) based on the orthogonal transform domain have been extensively investigated \cite{Zhang:15,Zhang:17,Bian:16,Yi:20}. Since they are based on a deterministic orthonormal basis, they can,in principle, achieve perfect reconstruction under full sampling and use their corresponding inverse transform to quickly reconstruct images, which significantly improves the reconstruction quality and reduces calculation time. In SPI, most patterns are realized using spatial light modulators. A digital micro-mirror device (DMD) is the most commonly used because of its advantages of high modulation speed, wide-band adaptation, non-polarization characteristics, and easy operation. The Fourier basis pattern is a gray-scale pattern, which slows down the modulation speed because the number of measurement matrices increases exponentially in the process of gray binarization and degrades image-reconstruction quality because of the inadequate quantization levels\cite{Zhang:17}. In contrast, the Hadamard patterns are already binarized to directly apply DMD to SPI, which simultaneously avoids the quantization noise and significantly reduces the modulation speed.

Several studies have recently been conducted on Hadamard basis ordering. Different sampling orderings affect the reconstruction quality when the Hadamard patterns are used for measurements at low sampling ratios. Therefore, images with better quality can be reconstructed with fewer measurements when the order of Hadamard patterns is rearranged, and the pattern that contributes most to the reconstruction results is measured first. Zhuoran \textit{et al.}\cite{Zhuoran:13} proposed Hadamard pattern order based on Walsh codes to optimize its performance. Sun \textit{et al.}\cite{SunRD:17} deeply investigated the Hadamard iteration law and proposed a new arrangement called the Russian Doll (RD order). Yu \textit{et al.}\cite{YuCC:19} used the concept of connected regions for matrix rearrangement and proposed a cake-cutting (CC order) strategy. The strategy of Xiao \textit{et al.}\cite{YuTV:20} does not need to count the connected regions but obtains a new Hadamard by sorting the normal-order Hadamard in the ascending order of the total variation (TV order). Vaz \textit{et al.}\cite{Vaz:20} compared five sorting methods, including Walsh order, CC order, Natural order, and Random order, among which CC order has the most outstanding performance. Xiao \textit{et al.}\cite{YuIE:20} added the total wavelet coefficients ascending order in the comparison and claimed that the comprehensive performance of the TV order and CC order is better than that of others. Vaz \textit{et al.}\cite{Vaz:22} presented two new orders, ascending scale and ascending inertia of the Hadamard basis, and they have good performance under a sampling ratio of $1\%-10\%$. Lopez-Garcia \textit{et al.}\cite{Lopez-Garcia:22} proposed an order based on the generalized basis vectors and zigzag fashion, which has at least the same performance and better algorithm complexity than the CC method.

In the past, when ordering Hadamard patterns, only the characteristics of the pattern were considered regardless of the location of the pattern on the spectrum. The Hadamard spectral energy distribution differs from the Fourier domain, and there are no such characteristics as central symmetry \cite{Wenwen:19} and radial correlation \cite{He:21}. The sampling paths of previous methods are usually simple shapes, such as circular, square, and zigzag \cite{Zhang:17}. In this study, we compare the spectral domains of the previous sorting methods and find that the existing orders can be approximated by sampling from low-frequency to high-frequency. Such sampling paths can reconstruct more reliable images at low-sampling rates. However, the imbalance of the two frequency components may result in indistinguishable details, blurred edges, and ringing effects.

This paper proposes two new sampling strategies for the Hadamard basis pattern. The core is to balance the weight of different spectral components to improve the image quality. Our selection method avoids the complicated process of analyzing each pattern and realizes the overall pattern selection. The energy of the Hadamard spectrum is mostly concentrated in the upper left corner of the spectrum. First, we found the direction (preliminary order (PO)) in which the energy is arranged from large to small on the Hadamard spectrum. Second, we used a probability function (PF) to randomly sample the spectral domain along this direction to obtain the Hadamard pattern. This approach is called the HSI imaging sampling strategy based on PO and PF (PO+PF). The computational complexity of generating Hadamard patterns in different orders has also been recently investigated \cite{Lopez-Garcia:22}. Sampling the spectrum directly is a good way to reduce the complexity of pattern generation. To improve the performance, we proposed an XY order to replace the PO. This method directly samples the Hadamard spectrum without analyzing the characteristics of each pattern or scanning the database. It has the advantages of low complexity and high generation speed. We compare PO+PF and XY+PF with the six previous Hadamard basis pattern ordering methods. The simulation and experimental results show that the proposed methods have a better reconstruction quality than the other six methods under the same sampling ratio, which can be used to obtain low-frequency information and a certain amount of high-frequency information about the image. In other words, the resolution of important details and edge structures is improved based on a reliable reconstruction of images. The improved XY method can generate orders of size 256 $\times$ 256 in only 0.445 s, which is three orders of magnitude shorter than that of the CC method. 

\section{Principles and Methods}

\subsection{Single-pixel imaging with Hadamard basis}
HSI is a SPI technique based on the Hadamard transform, and the Hadamard matrix is used as a pattern\cite{Geadah:1977}. A Hadamard matrix is an orthogonal matrix containing only 1 and -1, and its arbitrary rows and columns are orthogonal. The Hadamard matrix of order $2^k$ is defined as follows:
\begin{equation}
H_{2^{k}}=\left[\begin{array}{cc}
+H_{2^{k-1}} & +H_{2^{k-1}} \\
+H_{2^{k-1}} & -H_{2^{k-1}}
\end{array}\right], \quad H_{2}=\left[\begin{array}{cc}
+1, & +1 \\
+1, & -1
\end{array}\right] \quad H_{1}=[+1].
\end{equation}

\begin{figure}[htbp]
\centering
  \includegraphics[width=0.9\textwidth]{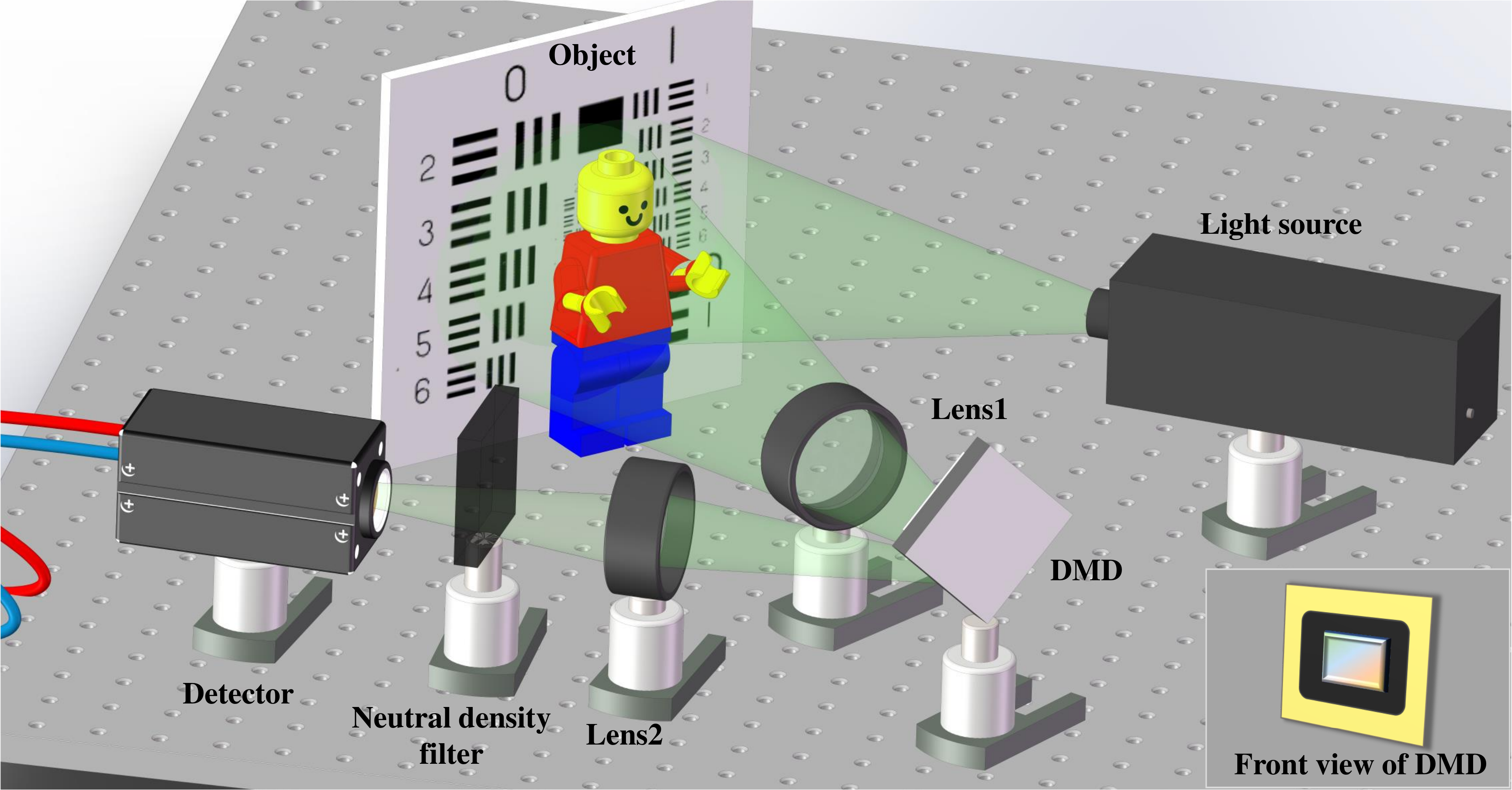}
  \caption{Schematic of experimental setup.}
\label{fig:fig1}
\end{figure}
As shown in Fig.~\ref{fig:fig1}, in the experiment, an object is illuminated by a light source and the reflected light from the object is collected by a lens and imaged on the spatial light modulator DMD. The modulated light is collected by a detector via a lens, which outputs the total light intensity. If the object $O(x,y)$ is of size $N\times N$, then we need to generate a Hadamard matrix $H_{{N^2}}$ and reshape the rows of this matrix into $N^2$ patterns $I^i(x,y) (i=1 \cdots N^{2})$ , each size is $N\times N$ pixels, where $x$ and $y$ represent the two-dimensional Cartesian position coordinates. Then, the patterns $I^i(x,y)$ are loaded on DMD to perform the modulation of the target object. Since DMD can only perform modulation of the binary matrix, the Hadamard matrix can be differentiated into a patterned pair $I_{+}^{i}\left( x,y \right)$ and $I_{-}^{i}\left( x,y \right)$ consisting of 0 and 1. $I_{+}^{i}\left( x,y \right)$ is generated by $\frac{1}{2}\left[ {{I}^{i}}\left( x,y \right)+1 \right]$, and the corresponding single-pixel detector measurement is $B_{+}^{i}$. $I_{-}^{i}\left( x,y \right)$ is generated by $\frac{1}{2}\left[ 1-{{I}^{i}}\left( x,y \right) \right]$, and the corresponding single-pixel detector measurement is $B_{-}^{i}$. Therefore, the differential HSI takes two measurements to acquire the measurement value $B^{i}$, 
\begin{equation}
B^{i} =B_{+}^{i}-B_{-}^{i} =\sum_{x, y} I^{i}(x, y) \cdot O(x, y), 
\end{equation}
which corresponds to the pattern $I^{i}(x,y)$. A measurement $B^{i}$ corresponds to a Hadamard spectral coefficient $h^{i}(u,v)$, where $u$ and $v$ represent the spectral parameters. The Hadamard spectrum $h(u,v)$ of the target image can be obtained after loading $N^2$ patterns sequentially. The original image can be obtained by performing the inverse Hadamard transform on the Hadamard spectrum. We used the TVAL3 \cite{Li:11} algorithm for reconstruction to reduce the reconstruction noise and improve the image quality. The above HSI principle can be expressed in CS as follows:
\begin{equation}
B=A \alpha.
\end{equation}
Here, $B$ is the measurement value, which is the column vector form of total measurements; $A$ consists of the selected rows in ${H_{{N^2}}}$, its size is $M \times N^{2}$, $M$ is the number of measurements required; $\alpha$ is the vectored form of $O(x,y)$.

\subsection{Hadamard pattern-selection strategy}
\begin{figure}[htbp]
\centering
  \includegraphics[width=0.9\textwidth]{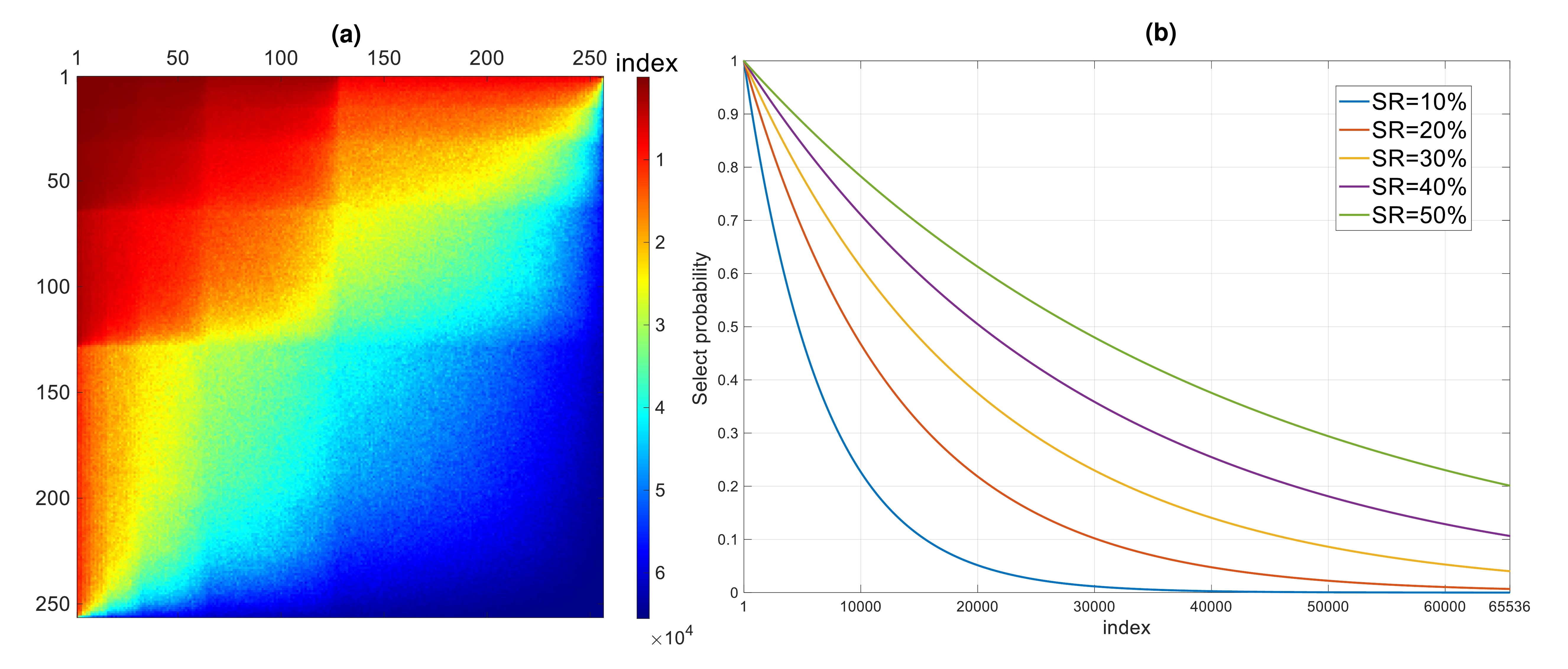}
  \caption{Hadamard pattern-selection strategy. (a) The sampling path of preliminary order (PO), $p = q = 256$. (b) Probability function graph at different sampling ratios. }
\label{fig:fig2}
\end{figure}

Unlike the previous methods that analyze the shape of the Hadamard patterns and determine the order, this paper uses the size of the coefficients on the Hadamard spectrum combined with the exponential function to determine the order in which Hadamard patterns are selected.

Our method is divided into two parts: The first step is to determine the preliminary order of the Hadamard spectral energy distribution from high to low using a dataset. We employ a commonly used image reconstruction dataset DIV2K\cite{Timofte:18,Qiu:21} in the simulation, which contains 800 images. Each image has been grayscaled and resized to $p\times q$ pixels, where $p\times q$ pixels is the size of the reconstructed image. We performed Hadamard transform on 800 images in the database and then normalized and summed all spectra to obtain the Hadamard power spectrum under this database. The coefficients in the power spectrum are sorted in a descending order to obtain the PO, which is represented by an index number. The larger the coefficient, the smaller the index number. Therefore, we found the direction in which the energy is arranged from large to small on the Hadamard spectrum. The index number has two dimensions (row and column), which locates its position on the Hadamard spectrum. The position determines which Hadamard pattern it corresponds to, so we can show the PO as a two-dimensional image (Fig.~\ref{fig:fig2}(a)).

The second step is to select the Hadamard patterns at the specified sampling ratio. In the Hadamard spectrum, the exponential function is used as a PF to randomly sample along the descending direction of the energy distribution, i.e., the direction of the PO. The probability function is given by:
\begin{equation}
E(n)=a^{[(n-1) / (p \times q)]} \quad(n=1 \cdots p\times q), 
\end{equation}
where $a$ is the calculated value $( 0<a<1 )$, when the size of the pattern is determined, the value "a" varies only with the sampling ratio (SR = $M/(p \times q)$), as shown in Fig.~\ref{fig:fig2}(b); $n$ is index number of the PO, $E(n)$ is the probability that the point number $n$ is selected. This PF is used for random sampling to complete the selection of spectral points. Accordingly, the Hadamard patterns corresponding to the spectral points are used to reconstruct the object.

\begin{figure}[htbp]
\centering
  \includegraphics[width=0.95\textwidth]{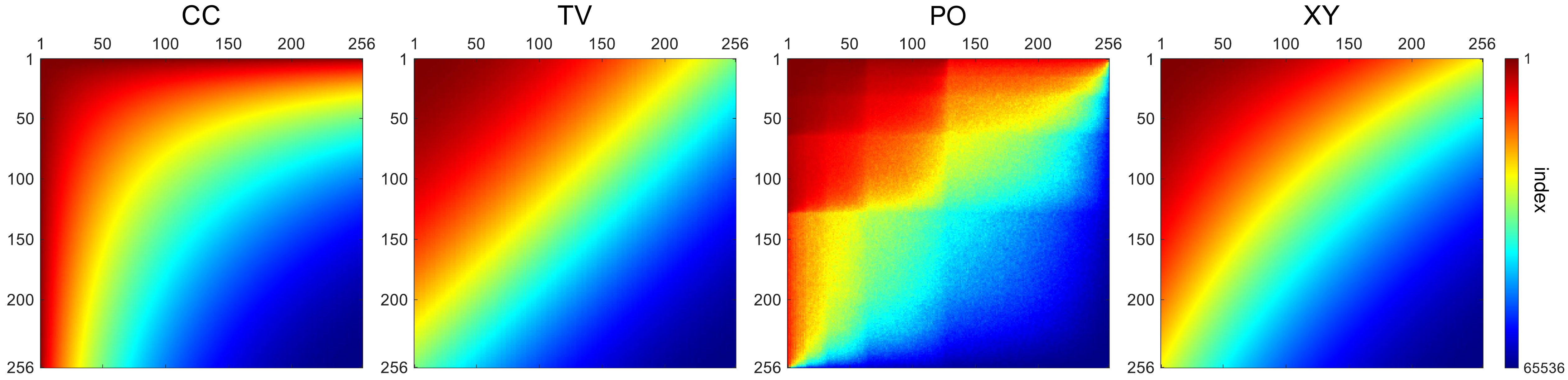}
  \caption{Sampling paths of the Hadamard spectrum for CC order, TV order, PO and XY order. ($p = q = 256$)}
\label{fig:fig3}
\end{figure}

In the above method, the first step of our method is to determine the PO of the Hadamard spectrum. This method of using the database to obtain the sampling path is inefficient. As shown in Fig.\ref{fig:fig3}, by fully observing the sampling path of different orders on the spectrum, we combined the sampling characteristics of the previous orders with better performance (CC order and TV order) and the PO obtained from the database. Then, we present a new Hadamard pattern ordering method called XY order, which can be used to replace the PO and improve performance.

In the XY order, we use the Cartesian coordinates to determine the importance of spectral coefficients and then sort the coefficients and basis patterns accordingly. The generation of the XY order is as follows. (1) Define the upper left corner of the Hadamard spectrum as the coordinate origin, and establish a two-dimensional x-y Cartesian coordinate system. Each spectrum coefficient has its corresponding x and y values. The weight value $m$ of each point in the spectrum is calculated according to the following formula:
\begin{equation}
m=x \cdot y+\frac{x^{2}+y^{2}}{4}.
\end{equation}
(2) Sort the basis patterns in the ascending order of their m values, represented by the index. The patterns with smaller indices are at the front of the XY order. The order is arbitrary, especially for the basis patterns with the same $m$ value. 

\begin{table}[ht!]   
\begin{center}   
\caption{Generation time of different Hadamard orders.}  
\label{table:1} \begin{tabular}{ccccc}   
\hline
{} & {} & {} & {Patterns size} & {} \\
\hline   
{} & {Methods} & {$64 \time 64$} & {$128 \time 128$} & {$256 \time 256$} \\   
\hline 
{} & {CC} & {1.999} & {23.730} & {430.785} \\
{Time ($s$)}& {TV} & {0.091} & {1.305} & {16.659} \\
{} & {XY} & {\textbf{0.038}} & {\textbf{0.115}} & {\textbf{0.445}} \\
\hline   
\end{tabular}   
\end{center}   
\end{table}

The complexity problem of different Hadamard orderings has been recently analyzed \cite{Lopez-Garcia:22}, and the generation time of the XY order was tested. Table \ref{table:1} shows the generation time required by the three sorting methods for different pattern sizes. Because the XY order can be generated only by using the spectral coordinates, it does not need to analyze the characteristics of patterns like other orders. Therefore, it has an extremely high generation speed and very low complexity. When generating a Hadamard pattern order of size 256 $\times$ 256, its speed is only 0.1$\%$ of the speed of CC order generation.

\begin{figure}[htbp]
\centering
  \includegraphics[width=0.8\textwidth]{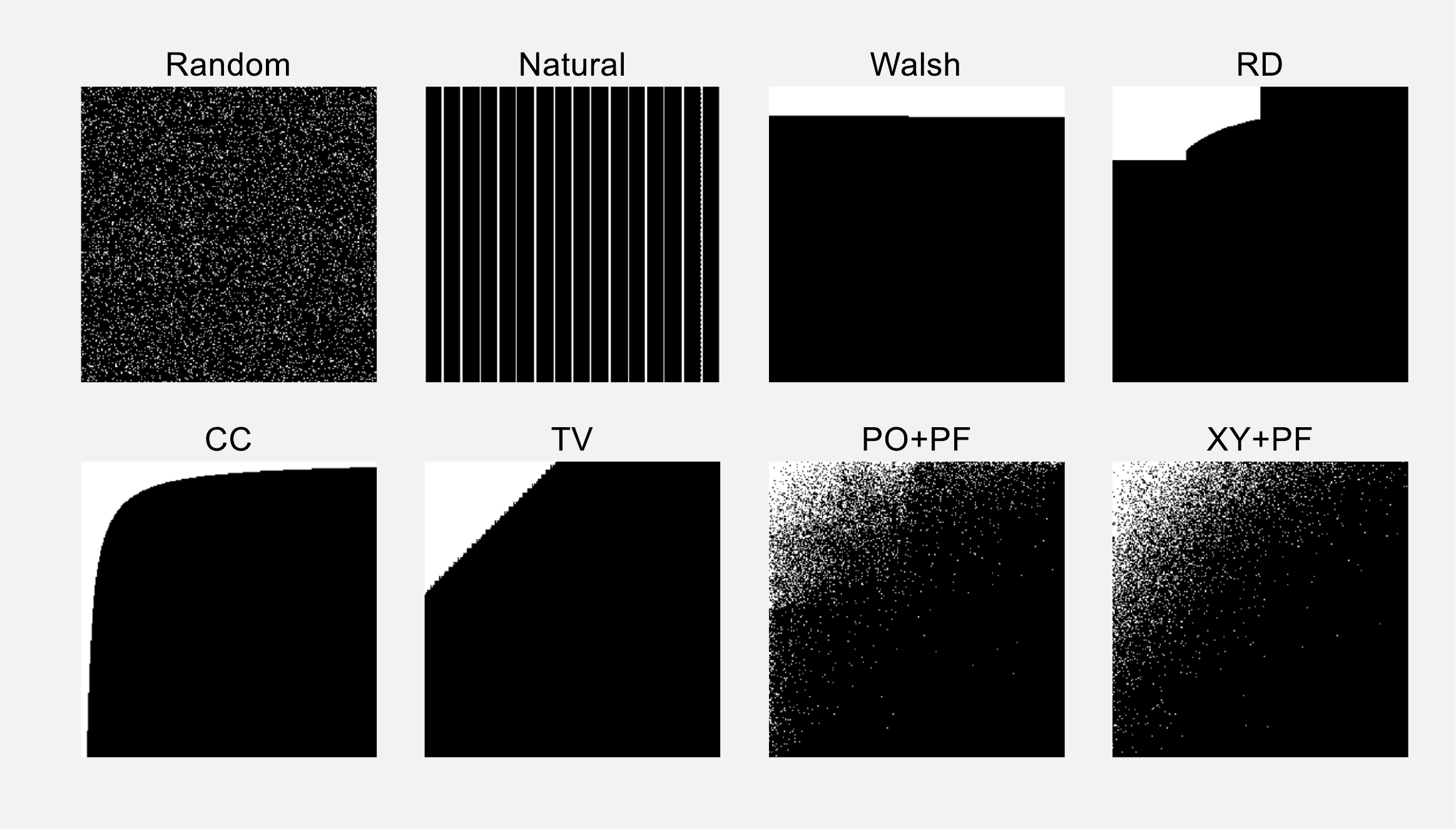}
  \caption{Sampling diagram in the Hadamard spectral domain for different sampling strategies.}
\label{fig:fig4}
\end{figure}

Fig.~\ref{fig:fig4} shows the sampling diagram in the Hadamard spectral domain for eight sampling strategies (Random order, Natural order, Walsh order, RD order, CC order, and TV order, PO+PF, XY+PF) when $p = q = 256$ under SR = $10\%$. The comparison methods include the classic and common Hadamard pattern sorting methods. In the spectral domain sampling diagram, the selected area is white, whereas the unselected area is black. At this time, $a$ is $4.5386\times 10^{-5}$, which is the calculated value under SR = $10\%$. All the calculations were performed on a desktop with an Intel Core i7-10700F CPU @ 2.90GHz and a RAM of 16 GB. This program was conducted in MATLAB R2021a. The sampling results in the spectral domain show that the orders with good comprehensive ability in the previous comparative work, such as the CC order and the TV order, tend to preferentially collect the low-frequency part of the spectrum with higher energy. Our method can sample high- and low-frequency parts of the spectrum proportionally. We also demonstrate the effect of the proposed method through experiments and simulations.

\section{Results}
\subsection{Assessment method}
To objectively evaluate the pros and cons of the sampling strategy, we selected two different indicators to quantitatively evaluate the image quality: peak signal-to-noise ratio (PSNR) and structural SIMilarity (SSIM). PSNR is the ratio of the maximum possible power of a signal to the destructive noise power that affects its representation accuracy. It is often used as an objective criterion for evaluating reconstruction quality in image reconstruction. It is defined as follows:
\begin{equation}
\begin{aligned}
P S N R&=10 \times \log _{10}\left(\frac{peakval^{2}}{M S E}\right), \\
M S E &=\frac{1}{p q} \sum_{x=1}^{p} \sum_{y=1}^{q}\left[O^{\prime}(x, y)-O(x, y)\right]^{2},
\end{aligned}
\end{equation}
where MSE represents the mean squared error of the reconstructed image from the original image; $p$ and $q$ are the length and width of the image; $O$ represents the original image signal; $O'$ represents the reconstructed image signal; $x$ and $y$ represent the row and column of the image; $peakval$ is either specified by the user or taken from the range of the image data type. The higher the PSNR, the less distortion of the reconstructed image, and the better the reconstruction quality.

SSIM is an index for measuring the similarity of two images. This method uses the mean to evaluate the brightness, the standard deviation to evaluate the contrast, the covariance to evaluate the structural similarity, and three different indicators are used to evaluate the reconstructed image.
\begin{equation}
\operatorname{SSIM}(x, y)=\frac{\left(2 \mu_{x} \mu_{y}+C_{1}\right)\left(2 \sigma_{x y}+C_{2}\right)}{\left(\mu_{x}^{2}+\mu_{y}^{2}+C_{1}\right)\left(\sigma_{x}^{2}+\sigma_{y}^{2}+C_{2}\right)} ,
\end{equation}
where $C_1$ and $C_2$ are constants; $\mu_{x}$,$\mu_{y}$, $\sigma_{x}$,$\sigma_{y}$,and $\sigma_{xy}$ are the mean, variance, and covariance of images x and y. The higher the SSIM, the better the similarity of the reconstructed image to the original image, and the better the reconstruction quality.

\subsection{Simulation results}
Two classic images were chosen, the USAF1951 resolution target and cameraman, as the target images for the simulation. Both of them have a resolution of 256 $\times$ 256 pixels and mixed 1$\%$ Gaussian noise to simulate the real situation.

\begin{figure}[htbp]
\centering
  \includegraphics[width=0.95\textwidth]{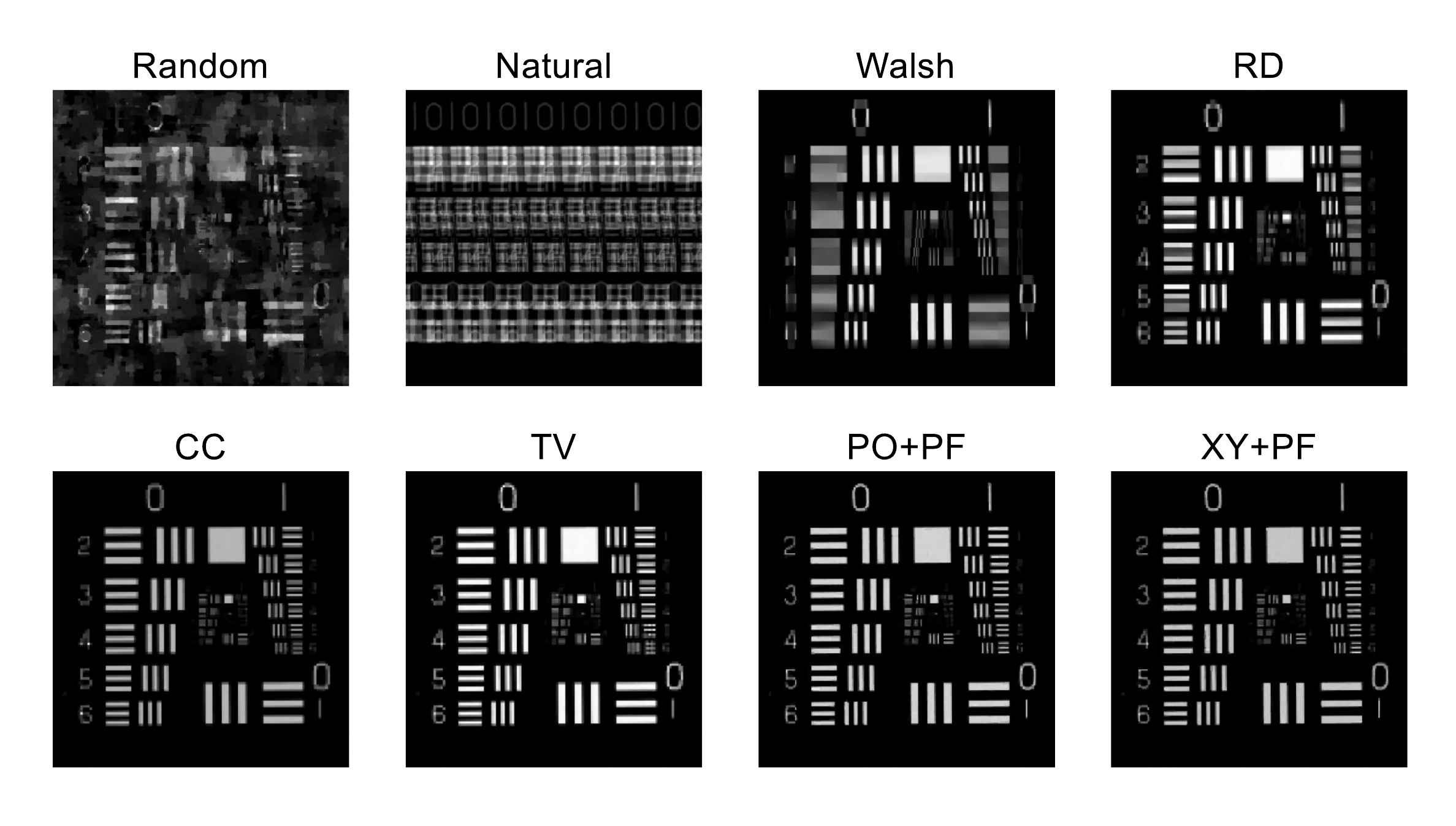}
  \caption{Resolution target simulation results under different sampling strategies. (Random order, Natural order, Walsh order, RD order, CC order, TV order, and our methods (PO+PF and XY+PF)) (SR = $10\%$)}
\label{fig:fig5} 
\end{figure}

\begin{figure}[htbp]
\centering
  \includegraphics[width=0.55\textwidth]{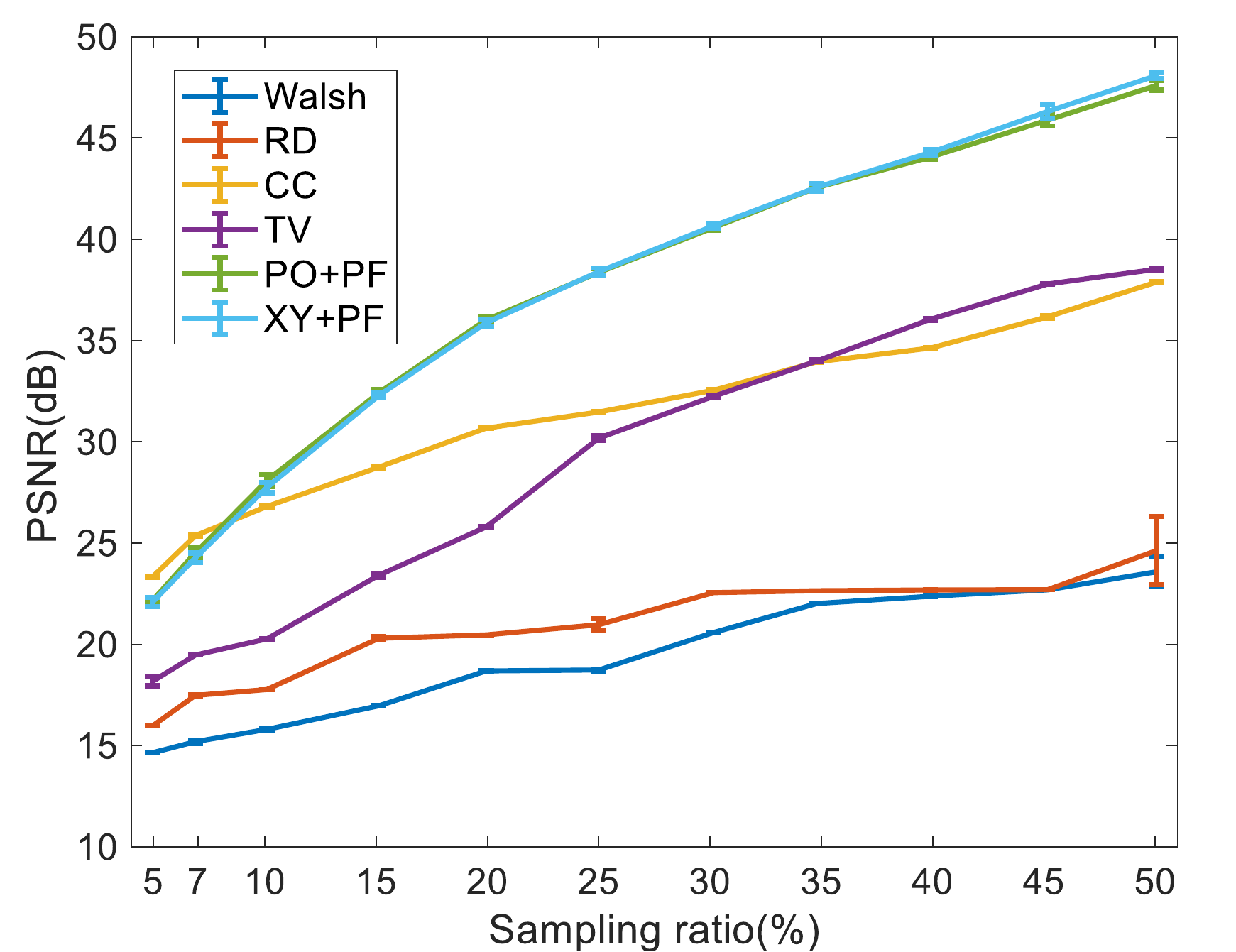}
  \caption{PSNR comparison of the overall effect of the resolution target.}
\label{fig:fig6}
\end{figure}

Fig.~\ref{fig:fig5} show the reconstruction results of the USAF1951 resolution target. Comparing the reconstruction results, the results of Random order and Natural order are extremely poor. It is almost impossible to distinguish the complete image. The results of the Walsh order are partially blurred, and the RD order, CC order, TV order, and our method recovered better images. The results obtained using our method have complete structures and clear details, better than those of other orders. To quantitatively compare the reconstruction effects of several sampling strategies, we used PSNR to assess overall imaging quality. As shown in Fig.~\ref{fig:fig5}, the reconstructed images of Random order and Natural order are significantly worse than the other order. Additionally, their PSNR values are only 13.057 and 11.5574 at SR = $5\%$, so they are excluded from the line chart for comparison. Fig.~\ref{fig:fig6} shows different sampling strategies using the TVAL3 algorithm to reconstruct the image, which is compared with the original image to obtain the PSNR. Each point shows the mean values of five measurements with added noise. Error bars have been added to all quantified data and the height of the error bars indicates the standard deviation. It can be seen that the overall effect of our methods is better than a series of previous methods when the sampling rate is above $10\%$. Although CC has better PSNR values at sampling ratios of $5\%$ and $7\%$, the difference between the CC order and our methods is not clearly discernible by visual observation.

\begin{figure}[htp]
\centering
  \includegraphics[width=0.9\textwidth]{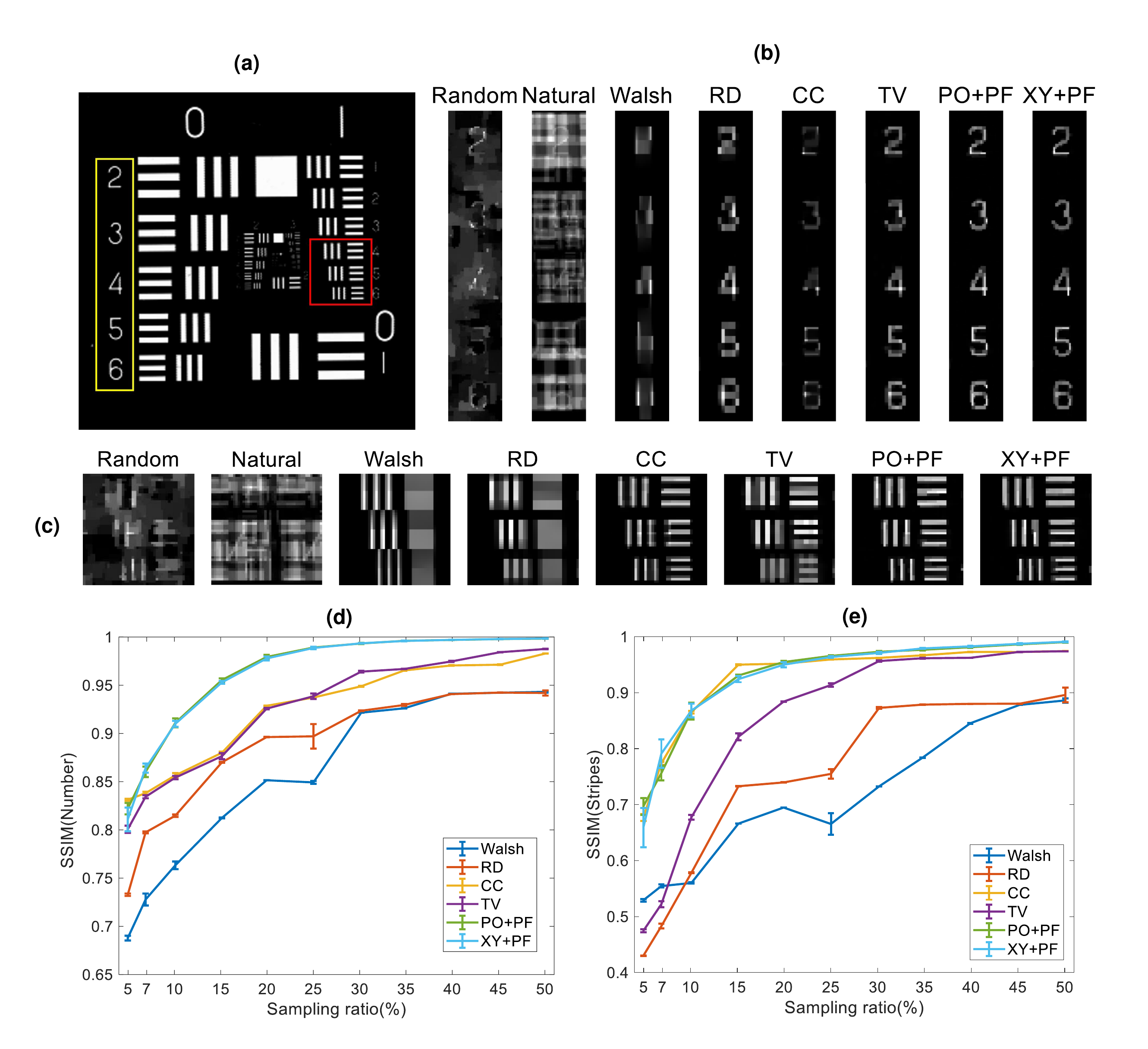}
  \caption{Detailed reconstruction results of the resolution target. (a) Two selected detailed areas of the resolution target. (b) The details of the number part (SR = $10\%$), (d) are the SSIM values in (b) area. (c) The details of the stripe part (SR = $10\%$), (e) are the SSIM values in (a) area.
}
\label{fig:fig7}
\end{figure}

We also selected detail regions to demonstrate that our method can improve the detail of the reconstructed image. Fig.~\ref{fig:fig7}(a) shows that the stripe part (red square) and number part (yellow square) in the image are selected for comparison experiments. In Figs.~\ref{fig:fig7} (b) and (c), for the stripe part, Random order, Natural order, Walsh order and RD order all have parts that cannot be reconstructed successfully. The stripe boundary of the CC order is blurred. The TV order did not have sufficient reconstruction ability for the thinnest fringes. Our methods recovered the sharpest stripes under the same sampling ratio. For the number part, it is obvious that the number of our methods is fine and the resolution is the highest. Figs.~\ref{fig:fig7} (d) and (e) show the results of the quantitative analysis of the detail part. The SSIM value of our method in the striped part is similar to that of the CC order, but it has much stronger performance than the other orders in the number part.

\begin{figure}[htbp]
\centering
  \includegraphics[width=0.9\textwidth]{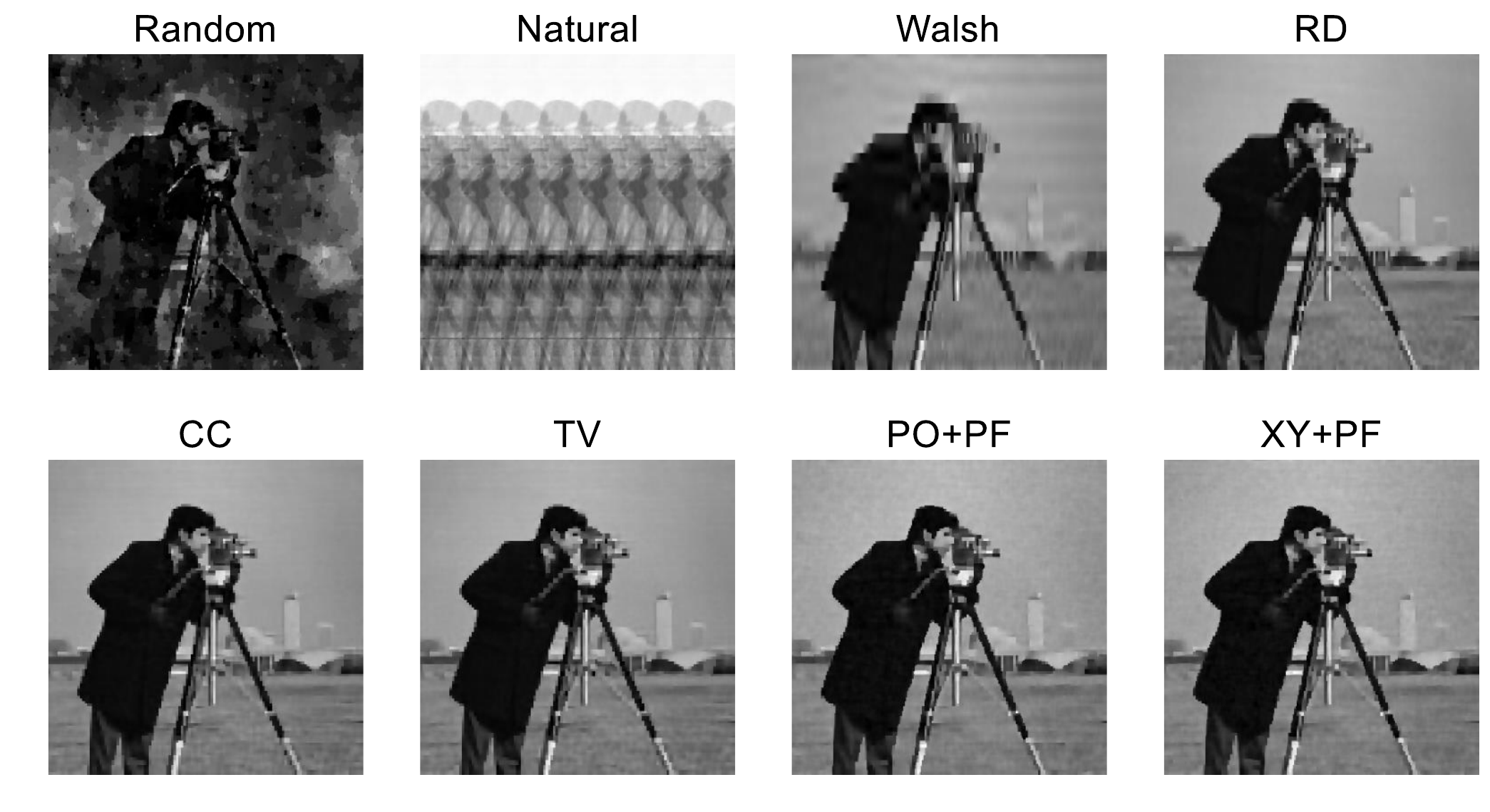}
  \caption{Cameraman simulation results of different sampling strategies. (Random order, Natural order, Walsh order, RD order, CC order, TV order, and our methods (PO+PF and XY+PF)) (SR = $10\%$)
}
\label{fig:fig8}
\end{figure}
\begin{figure}[htbp]
\centering
  \includegraphics[width=0.9\textwidth]{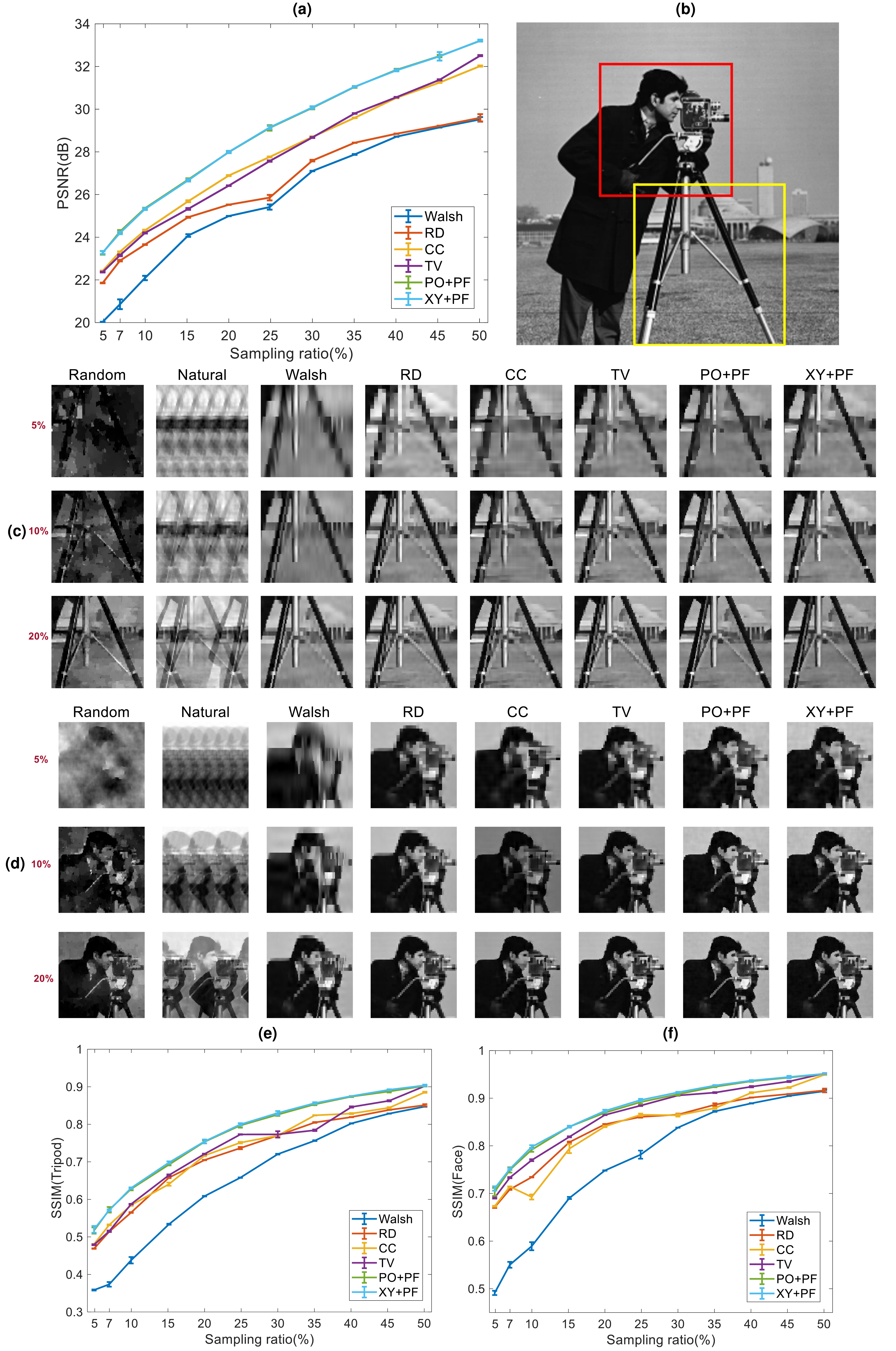}
  \caption{Overall and detailed reconstruction results of cameraman. (a) PSNR comparison of the overall effect of the cameraman. (b) Two selected detailed areas of the cameraman. (c) The details of the tripod part (SR = $10\%, 20\%, 30\%$), (e) are the SSIM values in (c) area. (d) The details of the face part (SR = $10\%, 20\%, 30\%$), (f) are the SSIM values in (c) area
}
\label{fig:fig9}
\end{figure}

A gray-scale image “cameraman” is further used for the test. Fig.~\ref{fig:fig8} shows that our method reconstructs many boundary details based on recovering the overall information. The quantitative comparison results of cameraman images are shown in Fig.~\ref{fig:fig9}(a). Considering the imaging quality of the overall image, our method is superior to the other six methods. Fig.~\ref{fig:fig9} shows the two comparison results of the details. As shown in Fig.~\ref{fig:fig9}(c), for the "tripod" part, our method can better outline the boundary contour of the main object and make the main object stand out from the background. For the “face” part in Fig.~\ref{fig:fig9}(d), the facial features of the reconstructed image obtained under the same conditions using our method are more prominent and clear. The ringing effect at the border of the hair and coat is suppressed, the jagged streaks of the hand-holding pole are reduced, and the detailed resolution is improved, indicating that our method outperforms the other six methods. The line graphs of SSIM in Figs.~\ref{fig:fig9}(e) and (f) well demonstrate the above analysis.

The comparison between the naked-eye observation and quantitative analysis confirmed that our method could distinguish the details well. It is worth noting that these two images do not belong to the DIV2K dataset, which can verify the universality of the proposed method.

\subsection{Experimental results}
The experimental device is shown in Fig.~\ref{fig:fig1}. The light source was an LED parallel light (KM-SPP505-W, beam size $\Phi$ 50 mm). The DMD (TI DLP7000) contained 1024 $\times$ 768 individually controllable micromirrors, and each micromirror was 13.68 $\mu$m $\times$ 13.68 $\mu$m. The loaded Hadamard pattern size was 128 $\times$ 128 pixels, and the DMD ran at a frame rate of 30 Hz, which is the upper limit of the detector's frame rate. The detector in this paper was a charge-coupled device (CCD, IMPERX B1620M), which was used only as a bucket detector\cite{Nie:22,Ferri:05}, and the spatial resolution of the CCD was not used. Each time the DMD was turned over, the CCD outputted the total light intensity value as the measurement value. Consequently, the collected measurement values were sent to a computer. The algorithm was used to recover the image. In this experiment, the target object was a toy figurine, and its background was the standard USAF1951 resolution target. PSNR and SSIM have been used for the quantitative analysis of experimental data. The reconstructed image obtained at a 100$\%$ sampling ratio was used as the original image.

\begin{figure}[htbp]
\centering
  \includegraphics[width=0.75\textwidth]{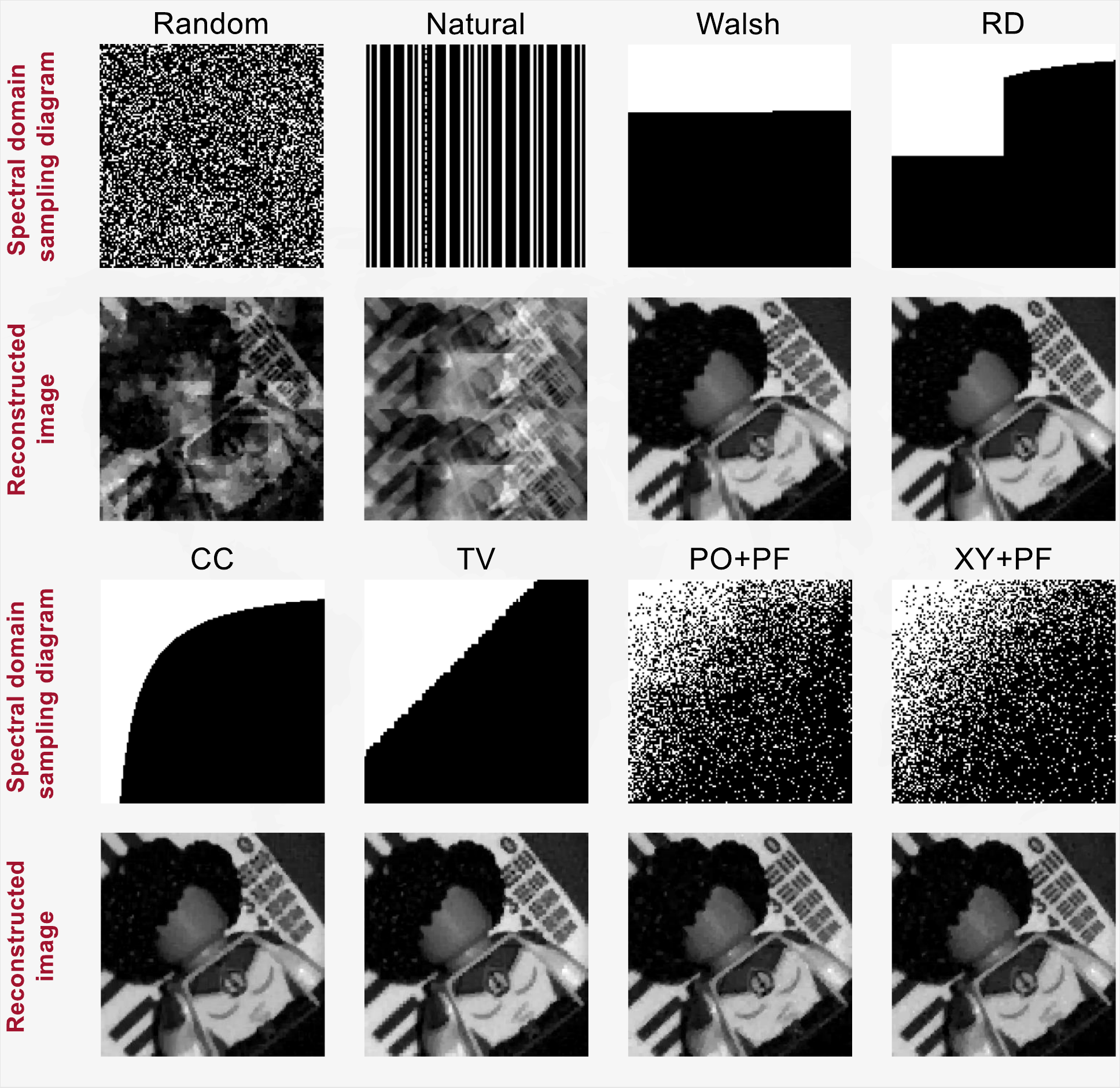}
  \caption{Experimental results of different sampling strategies. (Random order, Natural order, Walsh order, RD order, CC order, TV order, and our method (PO+PF and XY+PF))(SR= $30\%$)
}
\label{fig:fig10}
\end{figure}
\begin{figure}[htbp]
\centering
  \includegraphics[width=0.55\textwidth]{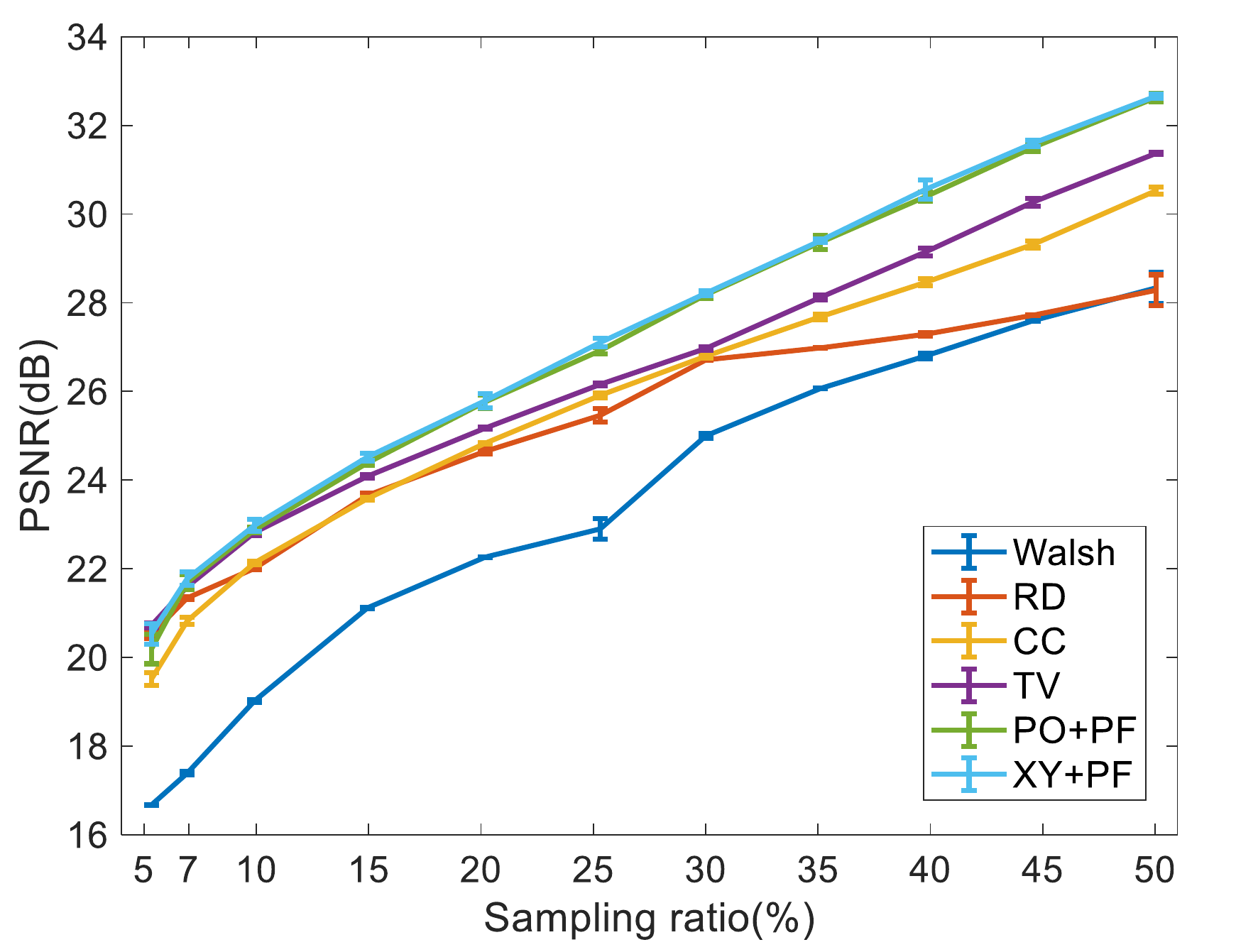}
  \caption{PSNR comparison of the overall effect of the experimental results.
}
\label{fig:fig11}
\end{figure}
\begin{figure}[htbp]
\centering
  \includegraphics[width=0.95\textwidth]{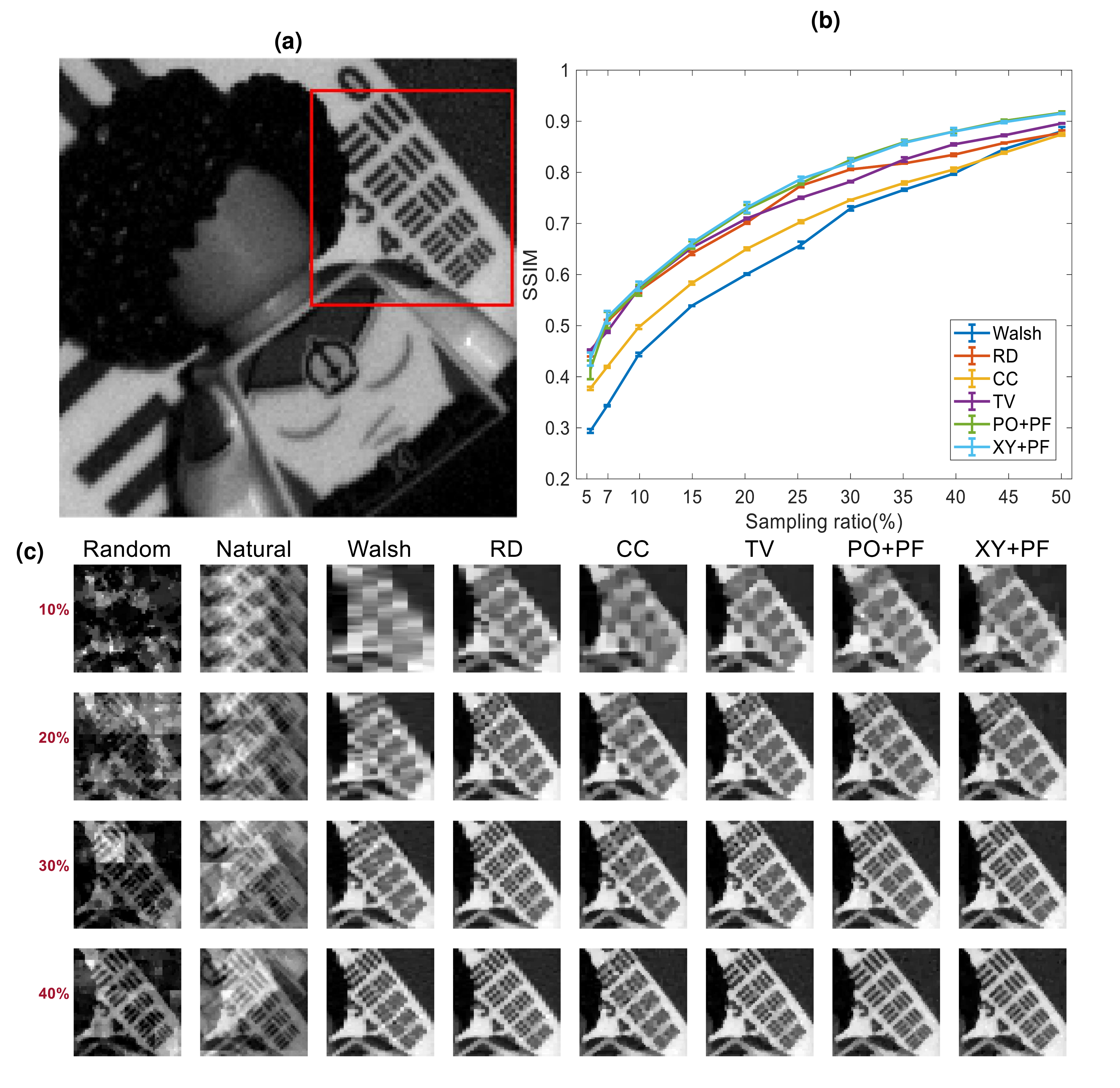}
  \caption{Detailed reconstruction results of the experiment. 
(a) Selected detailed area of the target object.
(c) The detail part (SR = $10\%, 20\%, 30\%, 40\%$), (b) is the SSIM value in (c) area.
}
\label{fig:fig12}
\end{figure}

Similarly, the experiment compared the reconstructions of Random order, Natural order, Walsh order, RD order, CC order, TV order, and our method. Fig.~\ref{fig:fig10} shows the experimental results of different sampling strategies with the same sampling ratio. Our methods reconstructed the gray-scale object (toy sculptures) well, and paid attention to the details of the resolution version in the background. Fig.~\ref{fig:fig11} shows the quantitative analysis of the overall reconstruction effect of the image. The proposed method has a better reconstruction quality than the other six methods. 

Fig.~\ref{fig:fig12} shows the results of the ability to distinguish fine details, which printed the indicated regions. The selected detailed areas are the Group 0, Element 2,3,4,5,6 portions of the USAF1951 resolution target. The mosaic blur appears at a low-sampling ratio when the Random order is applied. Images may be superimposed in the Natural order, and vertical fault stripes may appear in the Walsh order. For stripes, when the sampling ratio is 40$\%$, the fringe resolution of the RD order is low, the CC order fringes are cohesive and fuzzy, and the TV order can only distinguish thicker fringes. Meanwhile, the proposed method completely recovered the resolution stripes with the details, surpassing other sampling methods, such as TV order and CC order. Fig.~\ref{fig:fig12}(b) quantitatively confirms the superiority of this method.

The simulation and experimental results confirm that the proposed method can balance the acquisition of low-frequency and high-frequency information. The proposed method can also improve the detailed resolution based on a better reconstruction of the entire target image, which is consistent with the simulation results. Additionally, because the PF selection method retains a certain degree of randomness, it can show good performance for target objects other than the dataset, proving the universality of the method.

\section{Conclusion}
In this paper, we propose a detail-enhanced Hadamard basis sampling strategy. The prior knowledge of the high- and low-frequency information and PO can be obtained by measuring the database pictures. We used the exponential function is used as the PF to sample the PO to obtain the required Hadamard patterns. We also propose a XY order to replace the PO, which directly samples the Hadamard spectrum to select the pattern. The XY order exhibits good performance and extremely short generation time, generating a Hadamard patterns order of size 256 $\times$ 256 in only 0.445 s. The simulation and experimental results show that the imaging quality of the proposed method is better than that of the Random, Natural, Walsh, RD, CC, and TV orders. It can increase the resolution of important details, suppress the ringing effect, and highlight edge information. The proposed method can also have obvious effects on images outside the database and has good universality. It may be essential to study the characteristics of Hadamard spectral energy distribution, which has potential applications in adaptive selecting the proportion of high-frequency or low-frequency information in a reconstructed image.

\begin{backmatter}

\bmsection{Funding}
Beijing Institute of Technology Research Fund Program for Young Scholars 202122012.

\bmsection{Disclosures}
The authors declare no conflicts of interest.

\bmsection{Data availability}
Data underlying the results presented in this paper are not publicly available at this time but may be obtained from the authors upon reasonable request.

\end{backmatter}

\bibliography{sample}

\end{document}